\documentclass[aps,prd,showpacs,superscriptaddress,notitlepage,preprintnumbers]{revtex4-2}
\usepackage{graphicx,float,wrapfig,subfigure}
\usepackage{amsfonts,amsmath,amssymb,amstext}
\usepackage{latexsym}
 \usepackage{bm}
\usepackage{color}
\usepackage[normalem]{ulem}
\usepackage{hyperref}

\newcommand{\be}{\begin{equation}}
\newcommand{\ee}{\end{equation}}
\newcommand{\ba}{\begin{eqnarray}}
\newcommand{\ea}{\end{eqnarray}}

\definecolor{red}{rgb}{0.7,0,0}
\definecolor{green}{rgb}{0,0.5,0}

\begin{document}

\title{Photon and dilepton emission anisotropy for a magnetized quark-gluon plasma}
\date{March 8, 2024}

\author{Xinyang Wang}
\email{wangxy@aust.edu.cn}
\affiliation{Center for Fundamental Physics, School of Mechanics and Physics,
Anhui University of Science and Technology, Huainan, Anhui 232001, People's Republic of China}
\affiliation{Department of Physics, Jiangsu University, Zhenjiang 212013, People's Republic of China}
\affiliation{School of Fundamental Physics and Mathematical Sciences, Hangzhou Institute for Advanced Study, UCAS, Hangzhou 310024, People's Republic of China}

\author{Igor A. Shovkovy}
\email{igor.shovkovy@asu.edu}
\affiliation{School of Applied Sciences and Arts, Arizona State University, Mesa, Arizona 85212, USA}
\affiliation{Department of Physics, Arizona State University, Tempe, Arizona 85287, USA}

\begin{abstract}
We study the higher-order anisotropy coefficients $v_4$ and $v_6$  in the photon and dilepton emission from a hot magnetized quark-gluon plasma. Together with the earlier predictions for $v_2$, these results show a distinctive pattern of the anisotropy coefficients in several kinematic regimes. In the case of photon emission, nonzero coefficients $v_n$ (with even $n$)  have opposite signs at small and large values of the transverse momentum (i.e., $k_T\lesssim \sqrt{|eB|}$ and $k_T\gtrsim \sqrt{|eB|}$, respectively).  Additionally, the $v_n$ signs alternate with increasing $n$, and their approximate values decrease as $1/n^2$ in magnitude. The anisotropy of dilepton emission is well pronounced only at large transverse momenta and small invariant masses (i.e., when $k_T\gtrsim \sqrt{|eB|}$ and $M\lesssim \sqrt{|eB|}$). The corresponding $v_4$ and $v_6$ coefficients are of the same magnitude and show a similar alternating sign pattern with increasing $n$ as in the photon emission.
\end{abstract}
\maketitle

\section{Introduction}
\label{sec:introduction}

Quark-gluon plasma (QGP) is a state of extremely hot matter made of deconfined quarks and gluons that carry non-Abelian color charges \cite{Yagi:2005yb,Rischke:2003mt,Shuryak:2014zxa}. The existence of such a plasma state stems from the asymptotic freedom in quantum chromodynamics (QCD) \cite{Gross:1973id,Politzer:1973fx}. QGP was present naturally in the early Universe about a microsecond after the Big Bang. It can also be produced in heavy-ion collisions at the Relativistic Heavy Ion Collider (RHIC) in Brookhaven and the Large Hadron Collider (LHC) at CERN. The corresponding ``Little Bang" experiments allow one to study the fundamental properties of QGP \cite{Bzdak:2019pkr}.

Despite small sizes and short interaction times in relativistic collisions, experimental data provide a strong evidence that the QGP forms a strongly interacting viscous liquid \cite{PHOBOS:2004zne,PHENIX:2004vcz,STAR:2005gfr}. The flow measurements, quantified by the anisotropy coefficients $v_n$, support the scenario of QGP evolving hydrodynamically for a considerable fraction of its lifetime \cite{Heinz:2013th}. Theoretical models also indicate that the plasma has low viscosity \cite{Bernhard:2019bmu}, consistent with a strongly interacting regime. 

The dynamics responsible for the QGP production in heavy-ion collisions are complicated and only partially understood. One of the aspects in dire need of better understanding is the possible generation and evolution of background magnetic fields in noncentral collisions. Theoretical studies suggest that the initial magnetic field $B$ could be of the order of $m_\pi^2/e\approx 3\times 10^{18}~G$  \cite{Skokov:2009qp,Voronyuk:2011jd,Deng:2012pc,Bloczynski:2012en,Tuchin:2015oka,Guo:2019mgh}. Such an incredibly strong field could modify the thermodynamic and transport properties of QGP, trigger chiral anomalous effects (CME)~\cite{Fukushima:2008xe,Kharzeev:2007tn,Kharzeev:2007jp}, and ultimately affect numerous observables. For reviews, see Refs.~\cite{Tuchin:2013ie,Kharzeev:2015znc,Huang:2015oca,Miransky:2015ava}. 

To verify whether the QGP in noncentral collisions is magnetized and to estimate the strength of the magnetic field, one can try scrutinizing the most promising electromagnetic observables. It is reasonable to start by analyzing the photon \cite{Yee:2013qma,Tuchin:2014pka,Zakharov:2016mmc} and dilepton emission rates \cite{Tuchin:2013bda}. First, the magnetic field affects the corresponding rates already at leading order in coupling. Second, the photons and dileptons are clean probes of the QGP at early times. Indeed, owing to their long mean-free path, they do not suffer much from rescattering in a small volume of the plasma.

The heavy-ion experiments reveal that direct photons have a sizable elliptic flow, quantified by a large ellipticity coefficient $v_2$ \cite{Adare:2011zr,Adare:2015lcd,Acharya:2018bdy}. Their flow appears to be comparable to that of hadrons, which is truly surprising. Unlike hadrons, the direct photons are emitted at early times of QGP when collective flow may not have had the chance to form yet. This is known as the ``direct photon" puzzle. Many theoretical studies tried to address it \cite{Chatterjee:2005de,Schenke:2006yp,Chatterjee:2008tp,vanHees:2011vb,Linnyk:2013wma,Gale:2014dfa,Muller:2013ila,vanHees:2014ida,Monnai:2014kqa,Dion:2011pp,Liu:2012ax,Vujanovic:2014xva,McLerran:2014hza,McLerran:2015mda,Gelis:2004ep,Hidaka:2015ima,Linnyk:2015rco,Vovchenko:2016ijt,Koide:2016kpe,Turbide:2005bz,Tuchin:2012mf,Basar:2012bp}. In our detailed studies of the differential rates in Refs.~\cite{Wang:2020dsr,Wang:2021eud}, in particular, we argued that a large positive $v_2$ of the direct photons may be explained by the presence of a strong background magnetic field in the QGP. It is fair to note that further phenomenological investigations are needed to settle the issue. This study is one of the key steps in that direction. It extends the knowledge of the differential emission rates from a strongly magnetized plasma.

The dilepton emission is another complementary probe of the QGP. Since their spectra are not affected by the blueshift of the expanding medium, dileptons can serve as an excellent thermometer of the QGP~\cite{Rapp:2014hha}. On the other hand, the dilepton rate should be affected by the magnetic field \cite{Sadooghi:2016jyf,Bandyopadhyay:2016fyd,Bandyopadhyay:2017raf,Ghosh:2018xhh,Islam:2018sog,Das:2019nzv,Ghosh:2020xwp,Chaudhuri:2021skc,Das:2021fma}. Moreover, as we demonstrated in the earlier study \cite{Wang:2022jxx},  dilepton emission is characterized by a sizable ellipticity at small values of the invariant mass ($M\lesssim\sqrt{|eB|}$). In the same kinematic region, the rate is also strongly enhanced. It is fair to mention that the corresponding theoretical claims may be hard to verify systematically in current experiments. 

Here, we extend the previous studies by showing that the presence of a strong magnetic field in the QGP should be encoded not only in $v_2$, but also in high-order anisotropy coefficients. By using the same theoretical framework as in Refs.~\cite{Wang:2020dsr,Wang:2021eud,Wang:2022jxx}, here we obtain detailed theoretical predictions for the higher-order anisotropy coefficients $v_4$ and $v_6$ for a magnetized plasma at rest. Similarly to $v_2$, they show nontrivial dependence on the kinematic parameters. We argue that future detailed measurements of the photon and dilepton anisotropy coefficients could provide a distinctive fingerprint for verifying the presence of the background magnetic field in the plasma produced by noncentral heavy-ion collisions. Of course, to produce theoretical predictions for the corresponding heavy-ion observables, one needs to convolute the differential rates with the specific dynamical models of plasma. The latter task is left for future studies.

This paper is organized as follows. In Sec.~\ref{sec:model}, we introduce the key definitions and model assumptions in our study of the photon and dilepton emission from a hot magnetized QGP. The numerical results for higher-order anisotropy coefficients $v_4$ and $v_6$ are obtained and discussed in Sec.~\ref{sec:Numerical-results}. The summary of the main findings and conclusions are given in Sec.~\ref{sec:summary}. In  the Appendix, we quote the expression for the imaginary part of the Lorentz-contracted polarization tensor, which is needed for calculating the photon and dilepton rates.

\section{Model}
\label{sec:model}

Here we make the same model assumptions about the QGP as in Refs.~\cite{Wang:2020dsr,Wang:2021eud,Wang:2022jxx}. We consider a plasma made of the lightest up and down quarks. While the quantitative results may change slightly with the inclusion of the strange quarks, all qualitative results are to remain the same. For simplicity, we also assume that the masses of the up and down quarks are equal, i.e., $m_u = m_d =m = 5~\mbox{MeV}$. It is a good approximation for the QGP with a temperature of several hundred megaelectronvolts.

\begin{figure}[t]
\centering
  \subfigure[]{\includegraphics[width=0.375\textwidth]{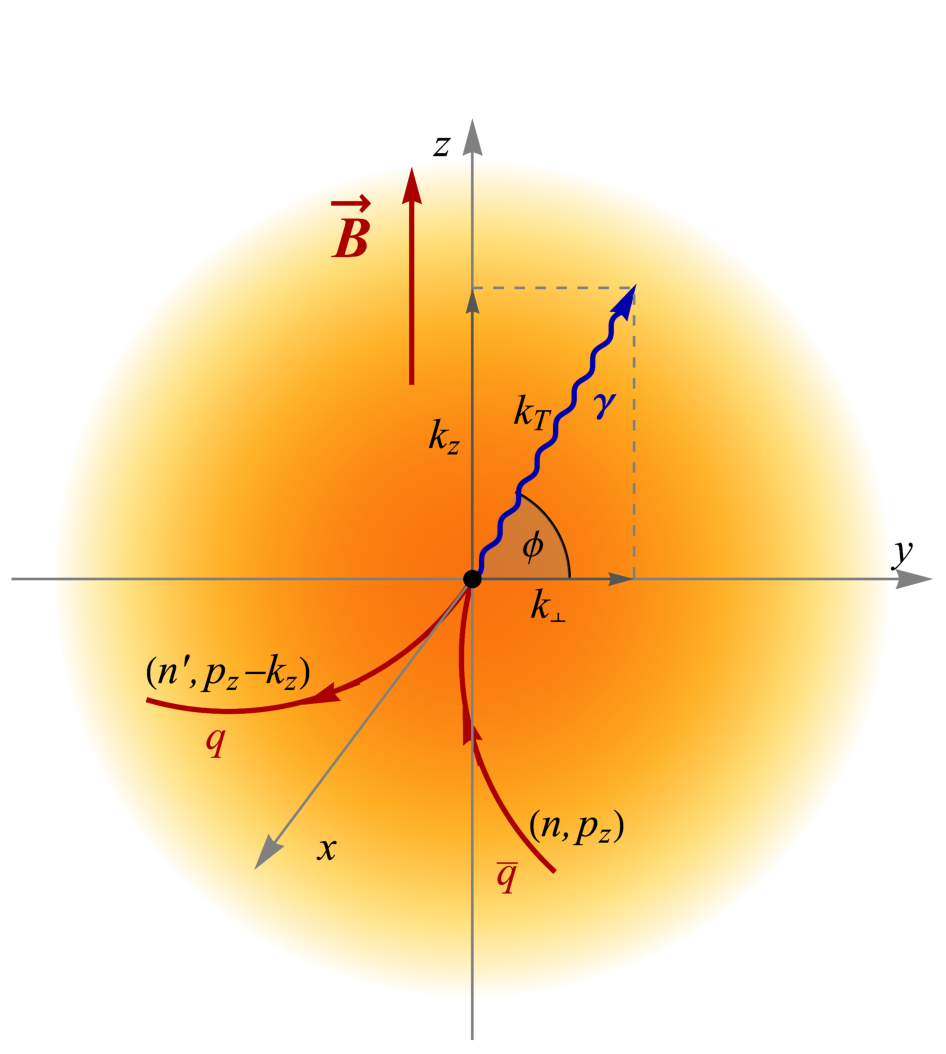}}
  \hspace{0.1\textwidth}
  \subfigure[]{\includegraphics[width=0.375\textwidth]{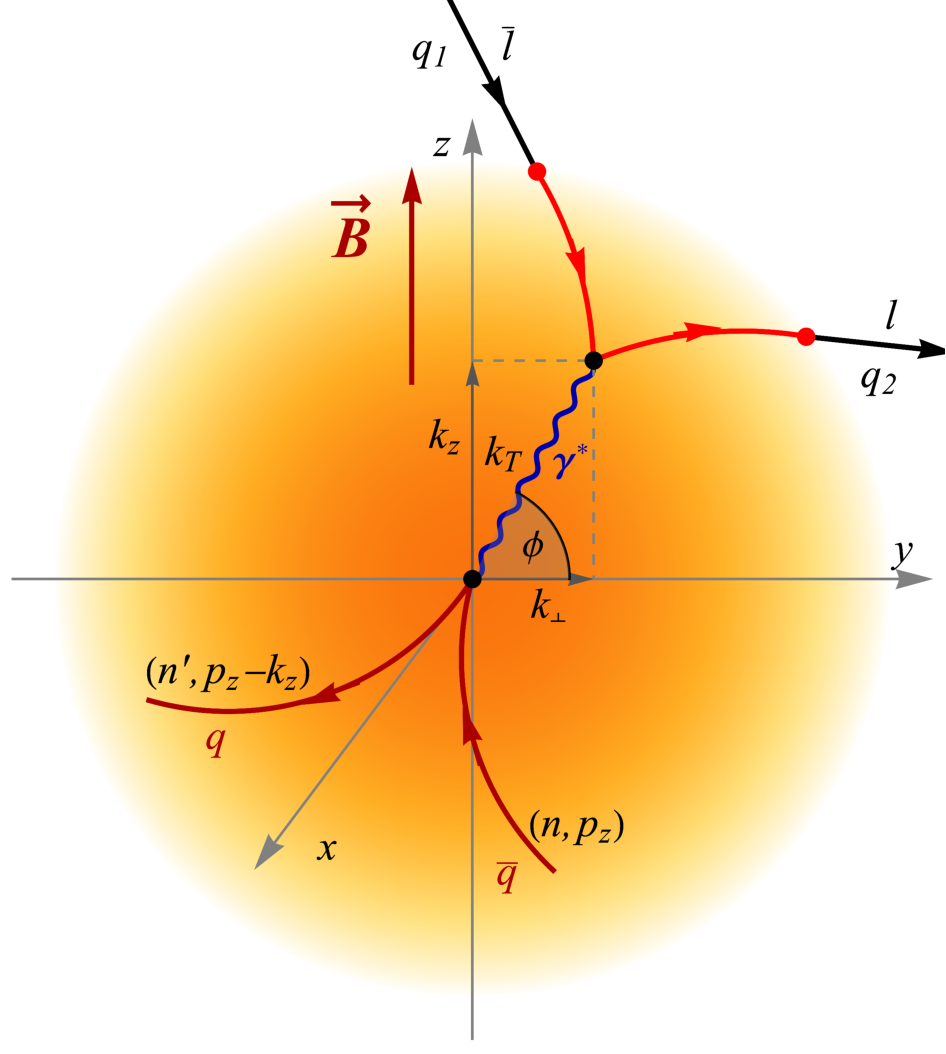}}
\caption{Schematic illustrations of the photon (a) and dilepton (b) emission from a magnetized plasma in the rest frame. The transverse momenta $k_T$ of the on-shell (a) and virtual (b) photons lie in the $y$-$z$ plane. The azimuthal angle $\phi$ is measured from the $y$ axis. The lepton momenta outside the magnetized plasma are $q_1$ and $q_2$ (b).}
\label{fig:illustrations}
\end{figure}

In this study, we consider the QGP plasma in the rest frame. By assumption, the magnetic field points along the $z$ axis. The two setups for photon and dilepton emission are illustrated schematically in the two panels in Fig.~\ref{fig:illustrations}.

In the case of photon emission, the corresponding four-momentum $k^\mu=(\Omega,\mathbf{k})$ satisfies the on-shell condition $k^2\equiv k_\mu k^\mu =0$. In the dilepton case, on the other hand, the photon $\gamma^{*}$ is virtual. Its momentum describes a lepton pair and satisfies a different on-shell condition, i.e., $k^2=M^2$, where $M$ is the dilepton invariant mass. Note that, without loss of generality, we can set $k_x = 0$ in the rest frame. The nonzero transverse components of the momentum are 
\begin{equation}
\label{kT-phi-parametrization}
k_y = k_T \cos(\phi),\qquad k_z =k_T \sin(\phi),
\end{equation}
where $k_T=\sqrt{k_y^2+k_z^2}$ is the magnitude of the transverse momentum and $\phi$ is the azimuthal angle measured from the $y$ axis. (The transverse component of the momentum $k_T$ should not be confused with $k_\perp=\sqrt{k_x^2+k_y^2}$, which is the component perpendicular to the magnetic field.)

\subsection{Photon emission rate and its anisotropy}

The thermal photon production rate from the QGP can be conveniently expressed in terms of the imaginary part of the retarded polarization tensor as follows \cite{Kapusta:2006pm}:
\begin{equation}
k^0\frac{d^3R}{dk_x dk_y dk_z}= \frac{d^3 R}{k_{T} d k_T d\phi dy} =-\frac{n_{B}(\Omega)}{(2\pi)^3} \mbox{Im}\left[\Pi^{\mu}_{R,\mu}\left(\Omega, \mathbf{k}\right)\right] ,
\label{diff-rate-photon}
\end{equation}
where $n_{B}(\Omega) = 1/\left[\exp\left(\Omega/T\right)-1\right]$ is the Bose-Einstein distribution function and $T$ is the temperature. The latter expression has the same form in QGP with and without a background field. However, a nonzero magnetic field can strongly affect the photon polarization tensor and, in turn, modify the photon emission rate. Below, we will utilize the leading-order one-loop expression for $\mbox{Im}\left[\Pi^{\mu}_{R,\mu}\left(\Omega, \mathbf{k}\right)\right]$ derived in Refs.~\cite{Wang:2020dsr,Wang:2021ebh,Wang:2022jxx} by using the Landau-level representation for quarks. For convenience, we also quote the corresponding result in the Appendix.  

When the differential rate (\ref{diff-rate-photon}) is known, the anisotropy coefficients $v_n$ can be evaluated as follows:
\begin{equation}
v_n(k_T) = \frac{1}{\mathcal{R}_0}  \int_0^{2\pi} \frac{d^3 R}{k_{T} d k_T d\phi dy}  \cos(n \phi) d \phi ,
\label{def:vn-photon}
\end{equation}
where the normalization factor is determined by integrating the emission rate over the azimuthal angle $\phi$, i.e.,
\begin{equation}
\mathcal{R}_0 = \frac{d^2 R}{k_{T} d k_T  dy} = \int_0^{2\pi} \frac{d^3 R}{k_{T} d k_T  dy d\phi} d \phi .
\end{equation}
We will use the definition in Eq.~(\ref{def:vn-photon}) to quantify the anisotropy of the photon emission from a hot magnetized QGP in Sec.~\ref{sec:Numerical-results}.

It is appropriate to comment on the approximation used here. When utilizing the one-loop polarization tensor in Eq.~(\ref{diff-rate-photon}), one accounts for the following three leading-order processes: (i) the quark splitting ($q\rightarrow q+\gamma $), (ii) the antiquark splitting ($\bar{q} \rightarrow \bar{q}+\gamma $), and (iii) the quark-antiquark annihilation ($q + \bar{q}\rightarrow \gamma$) \cite{Wang:2020dsr,Wang:2021ebh}. Their contributions  to the rate are of the order of $\alpha$, where $\alpha \equiv e^2/(4\pi)= 1/137$ is the fine structure constant. Recall that the same processes are forbidden by energy-momentum conservation in the absence of the magnetic field. Instead, leading contributions  at $B=0$ come from the gluon-mediated $2\to 2$ processes $q+g\rightarrow q+\gamma $, $\bar{q} +g \rightarrow \bar{q}+\gamma $, and $q + \bar{q}\rightarrow g+ \gamma$, where $g$ represents a gluon \cite{Kapusta:1991qp,Baier:1991em,Aurenche:1998nw,Steffen:2001pv,Arnold:2001ba,Arnold:2001ms,Ghiglieri:2013gia}. 
Formally, they are suppressed by an extra power of $ \alpha_s$, where $\alpha_s  \equiv g_s^2/(4\pi)$ is the QCD strong coupling constant. 

Unfortunately, the gluon-mediated $2\to 2$ processes have not been analyzed in a magnetic field. Thus, it is unclear how the relative contributions of the leading and subleading diagrams vary when one goes continuously from the zero-field to the strong-field limit. Here we will assume that the magnetic field is sufficiently strong for the leading-order contributions $\sim \alpha$ (from the $1\to 2$ and $2\to 1$ processes) to dominate the anisotropy coefficients. It can be true even in some cases when the subleading contributions $\sim \alpha\alpha_s$ (from the gluon-mediated $2\to 2$ processes) dominate the rates. With the current knowledge, however, we cannot establish a rigorous range of validity for the approximation used. It is an important issue and should be addressed in detail in future studies.

\subsection{Dilepton emission rate and its anisotropy}

Similarly to the photon emission, the differential dilepton production rate can be expressed in terms of the imaginary part of the photon polarization tensor, i.e., 
\begin{equation}
\frac{d R_{l\bar{l}}}{d^{4} k}=\frac{ \alpha}{12 \pi^4} \frac{n_{B}\left(\Omega\right)}{M^{2}} \mbox{Im}\left[\Pi^{\mu}_{R,\mu}\left(\Omega, \mathbf{k}\right)\right],
\label{rate_dilepton}
\end{equation}
where $n_B(\Omega) = (e^{\Omega/T}-1)^{-1}$ is the Bose-Einstein distribution function. Here we neglected the nonzero lepton masses and took into account that $k^2 \equiv \Omega^2 - k_\perp^2 -k_z^2=M^2$. Note that $\Omega =\sqrt{M^2+k_\perp^2 +k_z^2}$ and $d^{4} K = MdM k_T dk_T dy d\phi$.

To quantify the anisotropy of dilepton emission  in Sec.~\ref{sec:Numerical-results}, we will use the Fourier coefficients similar to those in Eq.~(\ref{def:vn-photon}), i.e., 
\begin{equation}
v_{n}(M, k_T) = \frac{\int_0^{2\pi} d\phi  \cos(n\phi) \left(d R_{l\bar{l}} /d^{4} k\right)}{\int_0^{2\pi} d\phi \left(d R_{l\bar{l}} /d^{4} k\right)} .
\label{def:vn-dilepton}
\end{equation}

It is instructive to emphasize that the approximation for the dilepton rate in Eq.~(\ref{rate_dilepton}), given in terms of the one-loop photon polarization tensor, is comparable to the leading-order result in the case of the vanishing magnetic field \cite{Cleymans:1986na}. Moreover, as shown in Ref.~\cite{Wang:2022jxx}, it reduces to the zero-field Born rate when the magnetic field goes to zero. Therefore, unlike the photon emission, the leading-order dilepton emission is under theoretical control in the whole range from the vanishing to strong magnetic fields.

\section{Results}
\label{sec:Numerical-results}

To extend our previous studies of the photon and dilepton emission rates in Refs.~\cite{Wang:2020dsr,Wang:2022jxx}, here we analyze the emission anisotropies in more detail. In particular, we study the higher-order coefficients $v_n$, as defined by Eqs.~(\ref{def:vn-photon}) and (\ref{def:vn-dilepton}).  Note that all odd coefficients $v_3$, $v_5$, etc. are vanishing in a magnetized plasma. It is the consequence of the rotation symmetry about the direction of the magnetic field. Here we will investigate the effect of the magnetic field on the high-order anisotropy coefficients $v_4$ and $v_6$. Note that the leading coefficient $v_2$, which measures the ellipticity of emission, was investigated in detail in Refs.~\cite{Wang:2020dsr,Wang:2022jxx}. By scrutinizing the angular dependence of the emission rates below, we will argue that such higher correlations hold interesting features that may become invaluable in quantifying the properties of the QGP produced in noncentral heavy-ion collisions. 

In numerical calculations, we express all mass and energy quantities in units of the (neutral) pion mass, $m_{\pi}\approx 0.135~\mbox{GeV}$. When presenting the results, however, we will display the transverse momenta and the dilepton invariant masses in gigaelectronvolts. To cover a substantial range of the parameter space without producing an overwhelming amount of data for the anisotropy coefficients, we will concentrate on the two representative choices of the magnetic field strength, $|eB|=m_\pi^2$ and $|eB|=5m_\pi^2$, and two representative values of temperature, $T=0.2~\mbox{GeV}$ and $T=0.35~\mbox{GeV}$, which are typical under the conditions in high-energy heavy-ion collisions \cite{PHENIX:2008uif,Wilde:2012wc,ALICE:2015xmh}.

As explained in Refs.~\cite{Wang:2020dsr,Wang:2022jxx}, the problem possesses a mirror symmetry with respect to the reflection in the reaction plane. Thus, the rates remain invariant when $\phi \to -\phi$. Taking into account also the parity symmetry ($y\to -y$), we see that the rate for the whole range of azimuthal angles from $\phi=-\pi$ and $\phi=\pi$ can be obtained from that in the range between $\phi=0$ and $\phi=\frac{\pi}{2}$. 

\subsection{Photon emission}
\label{vn-photon-emission}

Our earlier study in Ref.~\cite{Wang:2020dsr} showed that the photon emission from a magnetized hot QGP has a well-pronounced ellipticity characterized by a nonzero $v_2$. Moreover, its sign changes as some intermediate values of the transverse momentum. It is predominantly negative at small $k_T$ (i.e., $k_T\lesssim \sqrt{|eB|}$) and positive at  large $k_T$ (i.e., $k_T\gtrsim \sqrt{|eB|}$). Here we extend the study to the higher-order anisotropy coefficients $v_4$ and $v_6$. As we will see, they also deviate substantially from zero and show characteristic patterns of dependence on the transverse momentum. 

\begin{figure}[t]
\centering
\subfigure[]{\includegraphics[width=0.45\textwidth]{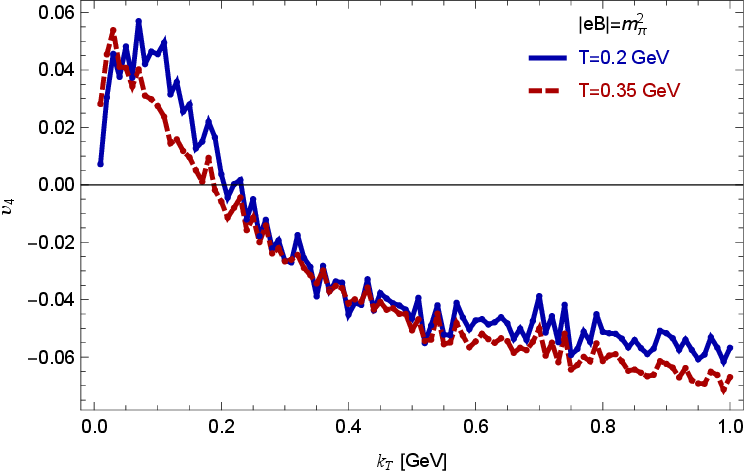}}
  \hspace{0.03\textwidth}
\subfigure[]{\includegraphics[width=0.45\textwidth]{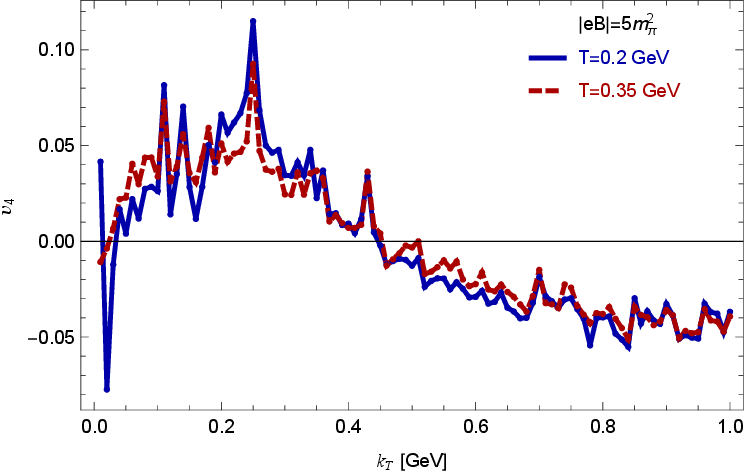}}
\caption{Anisotropic coefficient $v_4$ for the photon emission as a function of the transverse momentum $k_T$ 
for two different temperatures, $T=0.2~\mbox{GeV}$ (blue line) and $T=0.35~\mbox{GeV}$ (red line), and two different 
strengths of the magnetic field, $|eB|=m_\pi^2$ (a) and $|eB|=5m_\pi^2$ (b).}
\label{fig:photon_v4}
\end{figure}

The numerical results for $v_4$ in the photon emission are shown in Fig.~\ref{fig:photon_v4}. The left and right panels display the results for two different magnetic fields, $|eB|=m_\pi^2$ and $|eB|=5m_\pi^2$, respectively. In both cases, the blue solid and the red dashed lines correspond to two fixed temperatures, i.e., $T=0.2~\mbox{GeV}$ and $T=0.35~\mbox{GeV}$, respectively.  

We should note that the numerical data for $v_4$ (as well as other anisotropy coefficients below) appear to be very noisy, especially at small values of $k_T$. There are several reasons for such a behavior. In part, it is caused by the highly spiky dependence of the rates on the angular coordinate in the vicinity of the Landau-level thresholds, which are particularly strong at small $k_T$ \cite{Wang:2020dsr,Wang:2022jxx}. The jagged behavior is further exacerbated by a finite angular resolution of the numerical data. While some points happen to lie accidentally at or near sharp peaks, others fall near local minima.

The results reveal a clear qualitative pattern in the behavior of $v_4$ as a function of the transverse momentum. At relatively small momenta, $k_T\lesssim \sqrt{|eB|}$, $v_4$ tends to be positive. However, it becomes negative for 
$k_T\gtrsim \sqrt{|eB|}$. Notably, its absolute values are of the order of $0.05$. Such large $v_4$ values can be detectable in heavy-ion collisions if the background contributions due to other effects are under control.

\begin{figure}[t]
\centering
\subfigure[]{\includegraphics[width=0.45\textwidth]{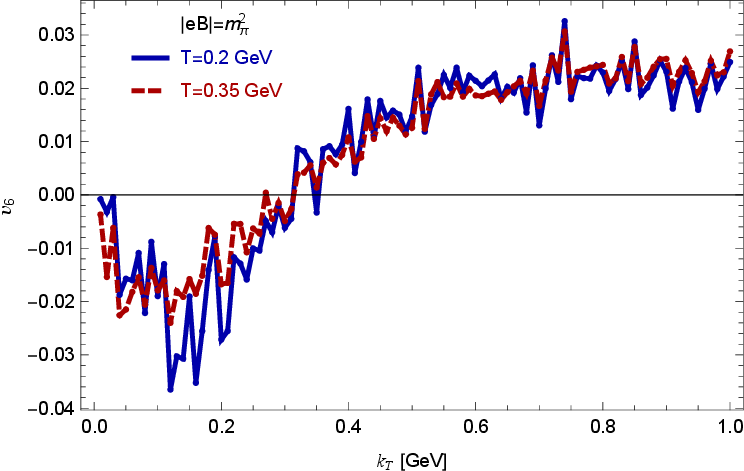}}
  \hspace{0.03\textwidth}
\subfigure[]{\includegraphics[width=0.45\textwidth]{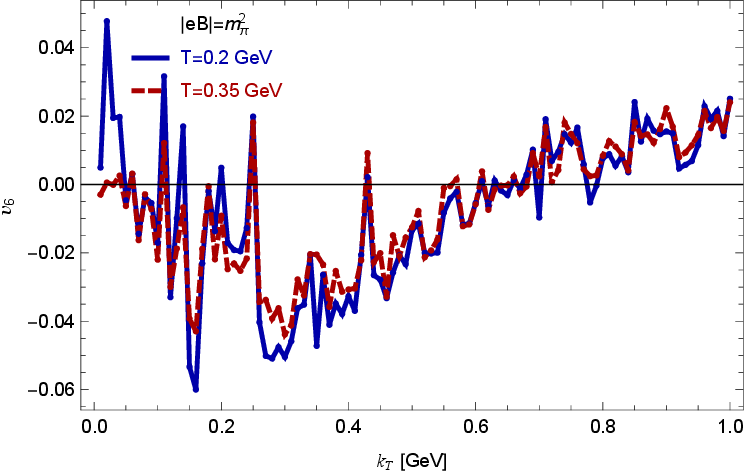}}
\caption{Anisotropic coefficient $v_6$ for the photon emission as a function of the transverse momentum $k_T$ 
for two different temperatures, $T=0.2~\mbox{GeV}$ (blue line) and $T=0.35~\mbox{GeV}$ (red line), and two different 
strengths of the magnetic field, $|eB|=m_\pi^2$ (a) and $|eB|=5m_\pi^2$ (b).}
\label{fig:photon_v6}
\end{figure}

The numerical results for $v_6$ are shown in Fig.~\ref{fig:photon_v6}. As before, the left and right panels display the results for two different magnetic fields, $|eB|=m_\pi^2$ and $|eB|=5m_\pi^2$, respectively. In both cases, the blue solid and the red dashed lines correspond to two fixed temperatures, i.e., $T=0.2~\mbox{GeV}$ and $T=0.35~\mbox{GeV}$, respectively. In all cases, the dependence of $v_6$  on the transverse momentum reveals similar qualitative features. It changes from a negative value at relatively small momenta, $k_T\lesssim \sqrt{|eB|}$, to positive values at relatively large momenta $k_T\gtrsim \sqrt{|eB|}$. The absolute values of $v_6$ are of the order of $0.02$, which are quite sizable too.  

The characteristic features of the photon anisotropy are summarized in Table~\ref{tab:anisotropy}. It is interesting to note the alternating signs of the anisotropy coefficients $v_{n}$ with increasing $n$. (Recall that all coefficients with odd $n$ vanish.) Another curious feature is the overall scaling of the magnitude, which goes as  $1/n^2$. The latter may be an approximate numerical result that holds only for the lowest three nonzero coefficients. However, tentatively it appears to remain true also for $v_{8}$, although the data become less reliable with increasing $n$ when the threshold effects from Landau levels produce many spikes in the angular dependence.

\begin{table}
\caption{\label{tab:anisotropy}Summary of nonvanishing photon and dilepton anisotropy coefficients $v_n$.}
\begin{ruledtabular}
\begin{tabular}{cccc}
 &\multicolumn{2}{c}{$v_n$ (photon emission)}&  $v_n$ (dilepton emission)\\ 
 \hline
  & $k_T\lesssim \sqrt{|eB|}$ & $k_T\gtrsim \sqrt{|eB|}$ & $k_T\gtrsim \sqrt{|eB|}$  \&  $M\lesssim \sqrt{|eB|}$\\ 
 \hline
 $v_2$ &  $\simeq -0.2$\footnotemark[1] &  $\simeq +0.2$\footnotemark[1] &  $\simeq +0.2$  \\
 $v_4$ &  $\simeq +0.05$ &  $\simeq -0.05$ &  $\simeq - 0.05$ \\
 $v_6$ &  $\simeq -0.02$ &  $\simeq +0.02$ &  $\simeq +0.02$ \\
 \end{tabular}
\end{ruledtabular}
\footnotetext[1]{From Ref.~\cite{Wang:2020dsr}.}
\end{table}

\subsection{Dilepton emission}
\label{vn-photon-emission}

As demonstrated in Ref.~\cite{Wang:2022jxx}, dilepton emission from a magnetized hot QGP shows a sizable ellipticity, described by a positive $v_2$ of the order of $0.2$, in the kinematic regime of small invariant masses (i.e., $M\lesssim \sqrt{|eB|}$) and large transverse momenta (i.e., $k_T\gtrsim \sqrt{|eB|}$). Here, we analyze the higher-order anisotropy coefficients $v_4$ and $v_6$. They also deviate noticeably from zero in the same kinematic regime. 

\begin{figure}[t]
\centering
{\includegraphics[width=0.45\textwidth]{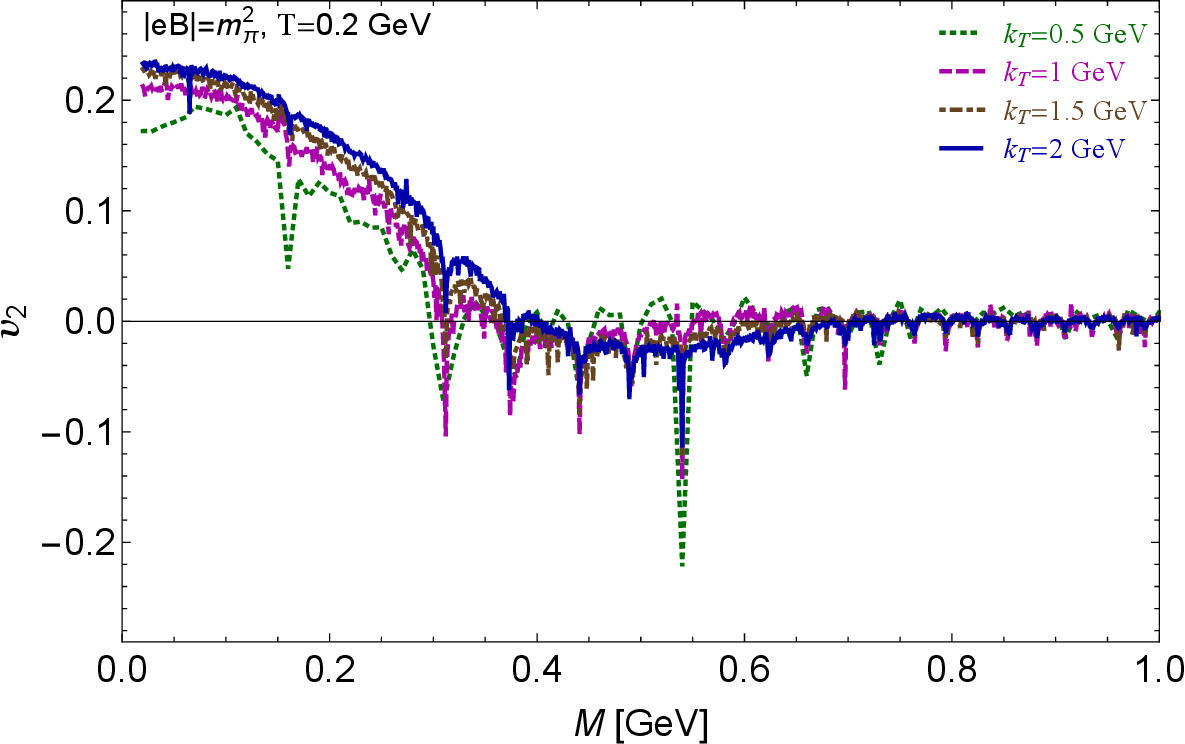}}
  \hspace{0.03\textwidth}
{\includegraphics[width=0.45\textwidth]{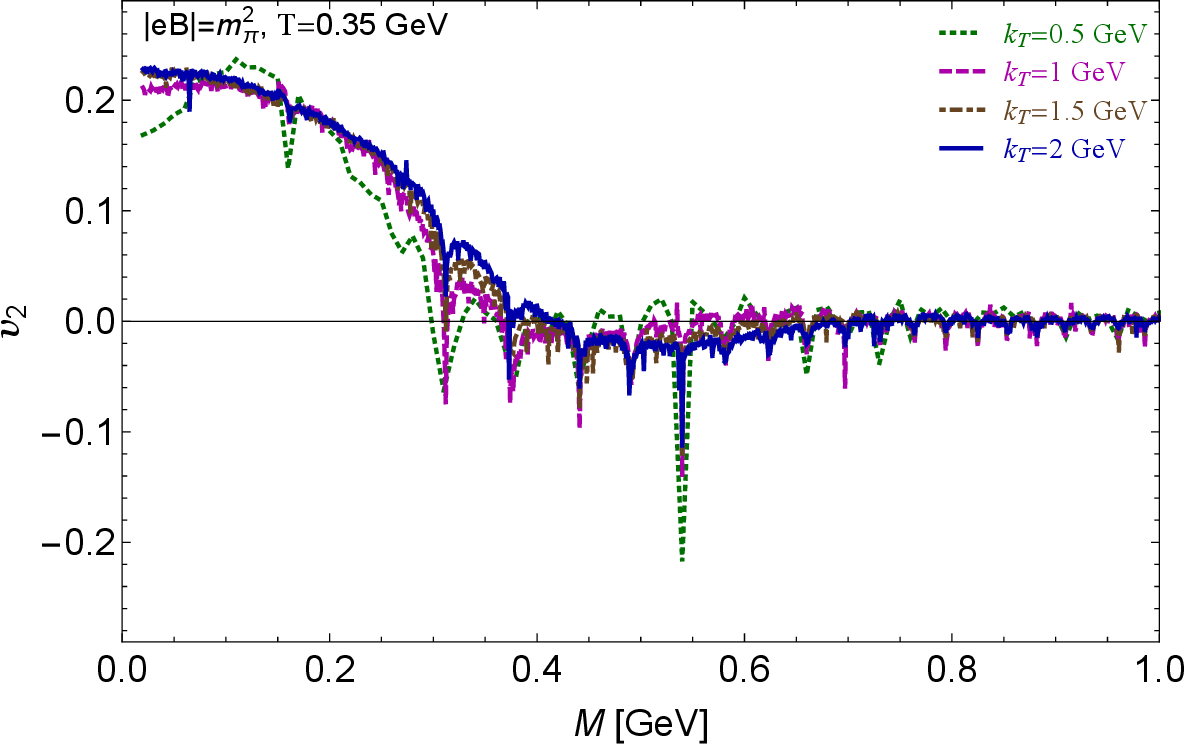}}\\[10pt]
{\includegraphics[width=0.45\textwidth]{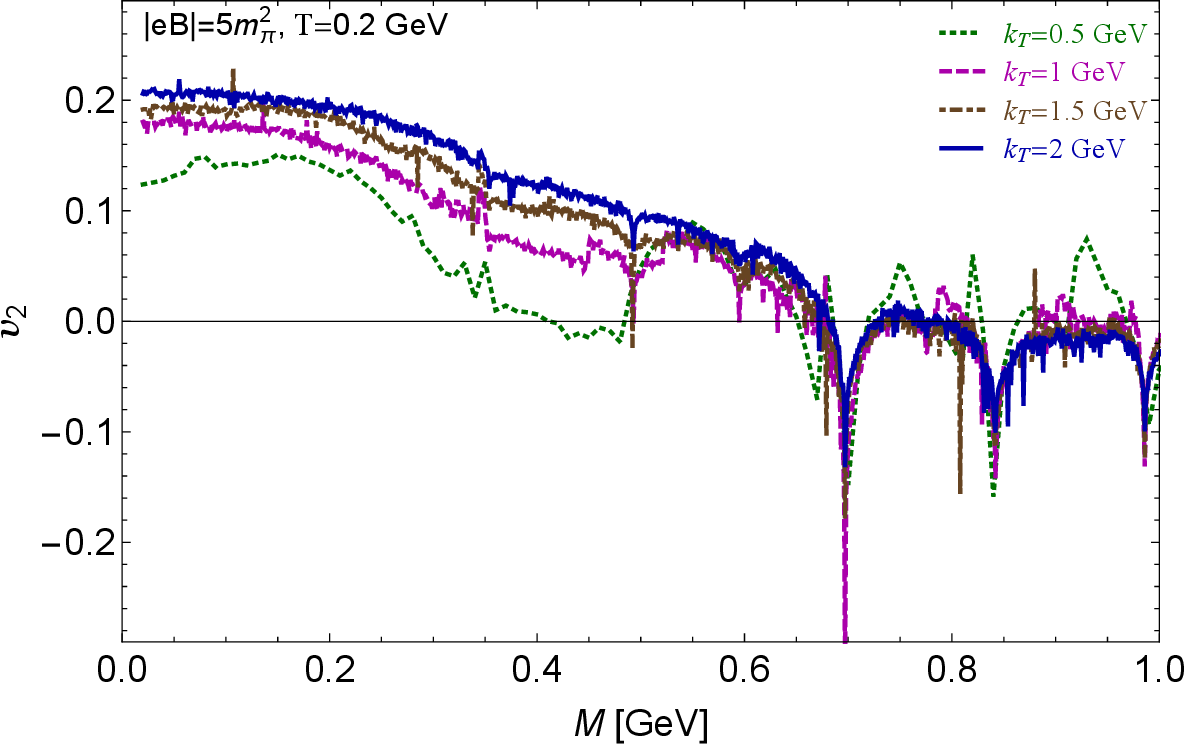}}
  \hspace{0.03\textwidth}
{\includegraphics[width=0.45\textwidth]{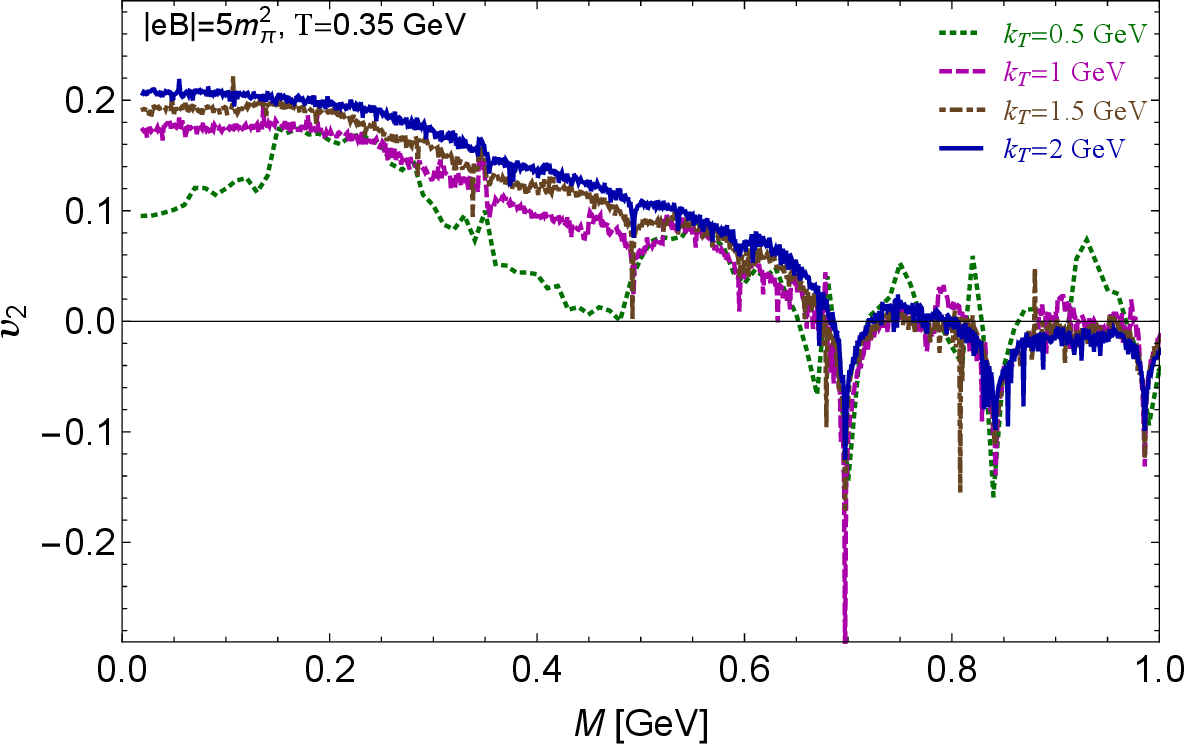}}\\
\caption{Anisotropic coefficient $v_2$ for the dilepton emission as a function of the invariant mass $M$ 
for several fixed values of the transverse momentum $k_T$. The top two panels correspond to $|eB|=m_\pi^2$ and the bottom two panels to $|eB|=5m_\pi^2$. The panels on the left are for $T=0.2~\mbox{GeV}$, and the ones on the right are for $T=0.35~\mbox{GeV}$. The data for $k_T=0.5~\mbox{GeV}$ are taken from Ref.~\cite{Wang:2022jxx}.}
\label{fig:dilepton_v2}
\end{figure}

Let us start by first reinforcing the results for the ellipticity of dilepton emission obtained in Ref.~\cite{Wang:2022jxx}. In particular, here we extend the previous calculations of the ellipticity coefficient $v_2$  to larger transverse momenta (up to $k_T = 2~\mbox{GeV}$) and increase the resolution in the invariant mass (i.e., from $\Delta M = 0.1~\mbox{GeV}$ down to $\Delta M = 0.01~\mbox{GeV}$). The corresponding new results are shown in Fig.~\ref{fig:dilepton_v2}. The four panels show the ellipticity coefficient $v_2$ as a function of the invariant mass for two temperatures, i.e., $T=0.2~\mbox{GeV}$ (two left panels) and $T=0.35~\mbox{GeV}$ (two right panels), and two magnetic fields $|eB|=m_\pi^2$ (two top panels) and $|eB|=5m_\pi^2$ (two bottom panels). For reference, we also included one of the older low-resolution datasets for $k_T=0.5~\mbox{GeV}$ from Ref.~\cite{Wang:2022jxx}. 

By comparing the dependence of $v_2$ on the invariant mass $M$ with the results in Ref.~\cite{Wang:2022jxx}, we find that the earlier conclusions are not only valid, but they also become more robust with the increasing of the transverse momentum. Furthermore, the current high-resolution data reconfirm that the ellipticity coefficient $v_2$ takes generically large positive values ($v_2\sim 0.2$) in the region of small invariant masses  (i.e., $M\lesssim \sqrt{|eB|}$). Its magnitude is comparable to the photon $v_2$ calculated in Ref.~\cite{Wang:2020dsr}. By comparing the data for the two different temperatures in Fig.~\ref{fig:dilepton_v2}, we also see that the temperature dependence of the dilepton $v_2$ is nearly negligible. For the large transverse momenta considered, of course, it should not be surprising. As we will see below, both $v_4$ and $v_6$ reveal a similarly weak temperature dependence.

\begin{figure}[t]
\centering
{\includegraphics[width=0.45\textwidth]{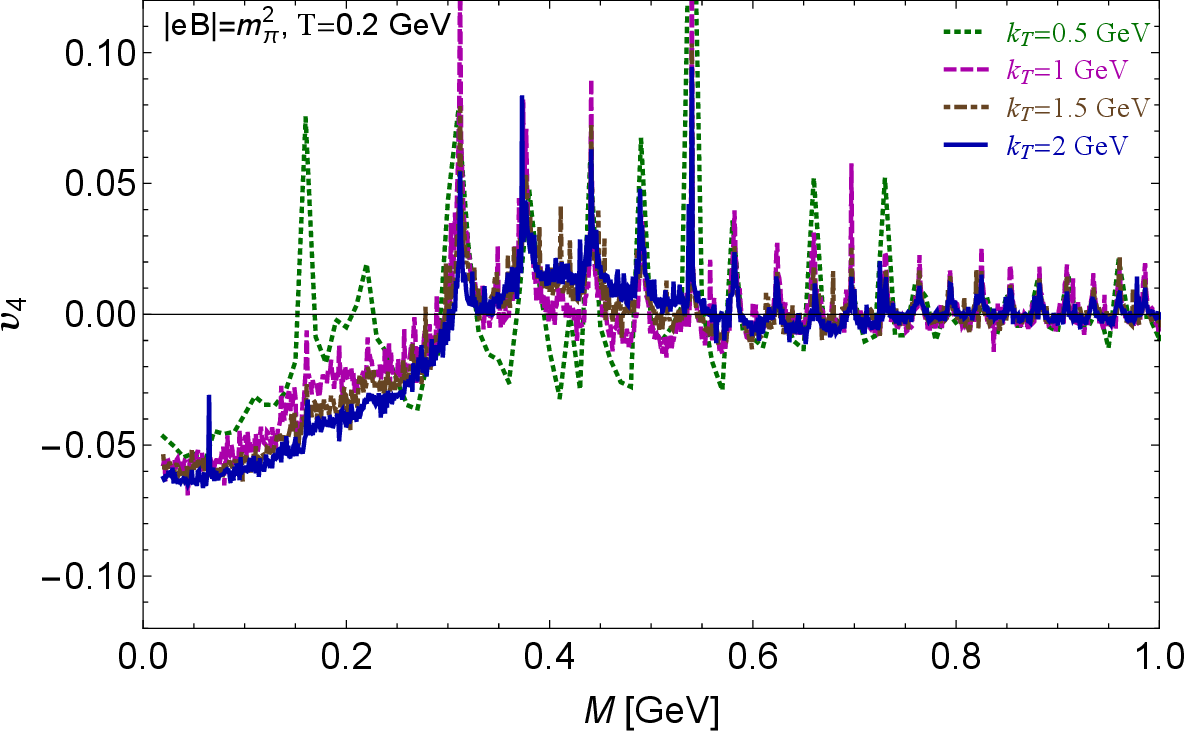}}
  \hspace{0.03\textwidth}
{\includegraphics[width=0.45\textwidth]{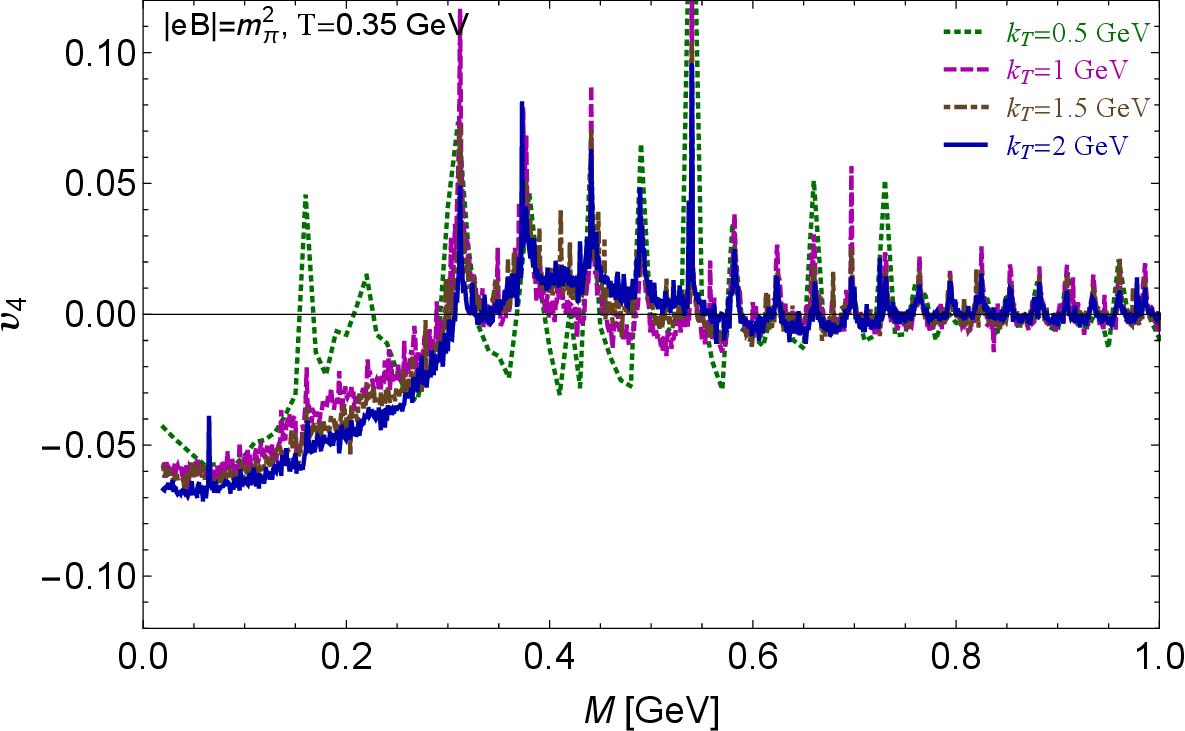}}\\[10pt]
{\includegraphics[width=0.45\textwidth]{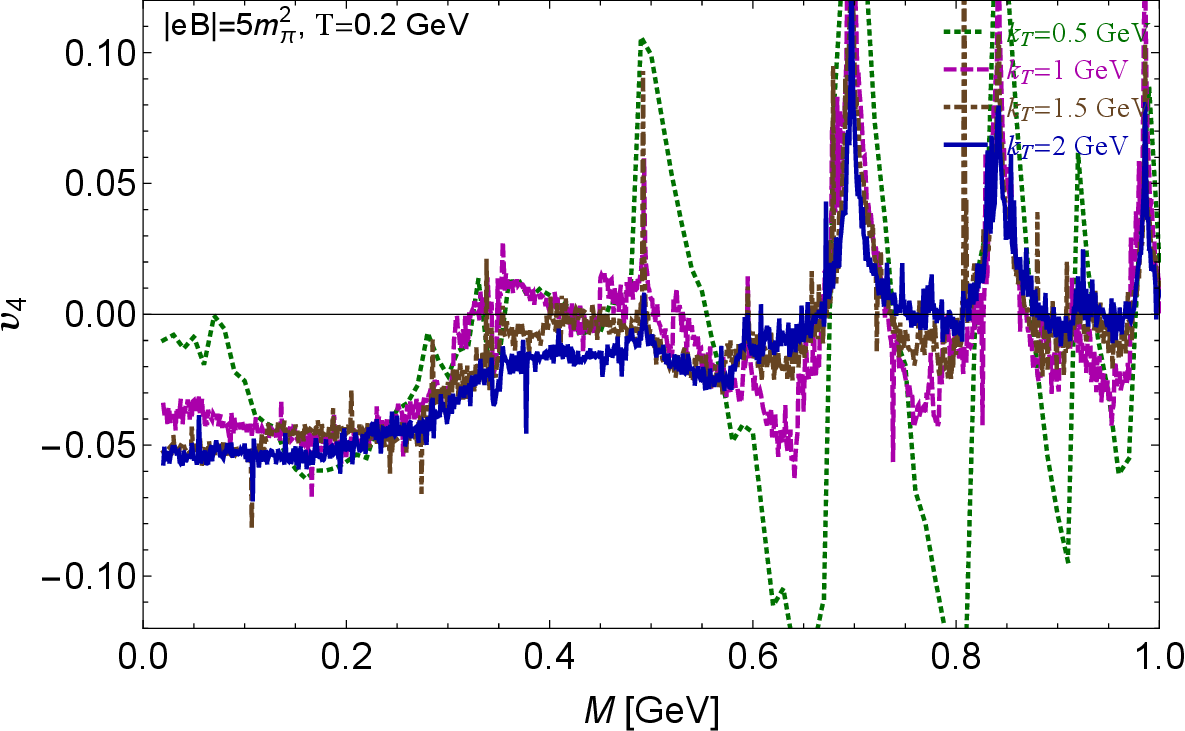}}
  \hspace{0.03\textwidth}
{\includegraphics[width=0.45\textwidth]{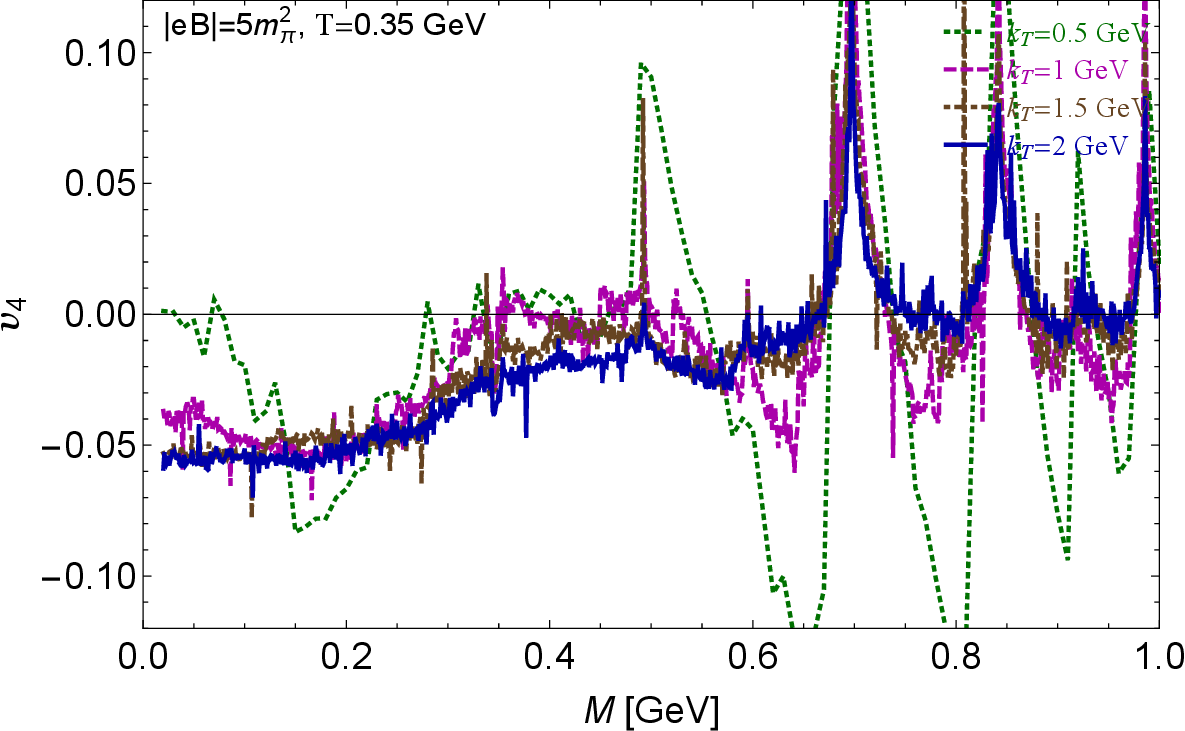}}\\
\caption{Anisotropic coefficient $v_4$ for the dilepton emission as a function of the invariant mass $M$ 
for several fixed values of the transverse momentum $k_T$. The top two panels correspond to $|eB|=m_\pi^2$ and the bottom two panels to $|eB|=5m_\pi^2$. The panels on the left are for $T=0.2~\mbox{GeV}$, and the ones on the right are for $T=0.35~\mbox{GeV}$.}
\label{fig:dilepton_v4}
\end{figure}

Now, let us turn to the higher-order anisotropy coefficients $v_4$ and $v_6$. We will concentrate our attention on the same kinematic region of small invariant masses and large transverse momenta, where the anisotropy is pronounced the most. The numerical results for the dilepton $v_4$ as a function of the invariant mass are shown in Fig.~\ref{fig:dilepton_v4}. As before, the four panels present the results for two temperatures, i.e., $T=0.2~\mbox{GeV}$ (two left panels) and $T=0.35~\mbox{GeV}$ (two right panels), and two magnetic fields $|eB|=m_\pi^2$ (two top panels) and $|eB|=5m_\pi^2$ (two bottom panels). As we see from Fig.~\ref{fig:dilepton_v4}, at small invariant masses, the coefficient $v_4$ tends to be negative with the absolute values of about $0.05$. Note that these are sizable by any reasonable standards. They are also comparable to the $v_4$ values in the photon emission at large transverse momenta, see Fig.~\ref{fig:photon_v4}. 

\begin{figure}[t]
\centering
{\includegraphics[width=0.45\textwidth]{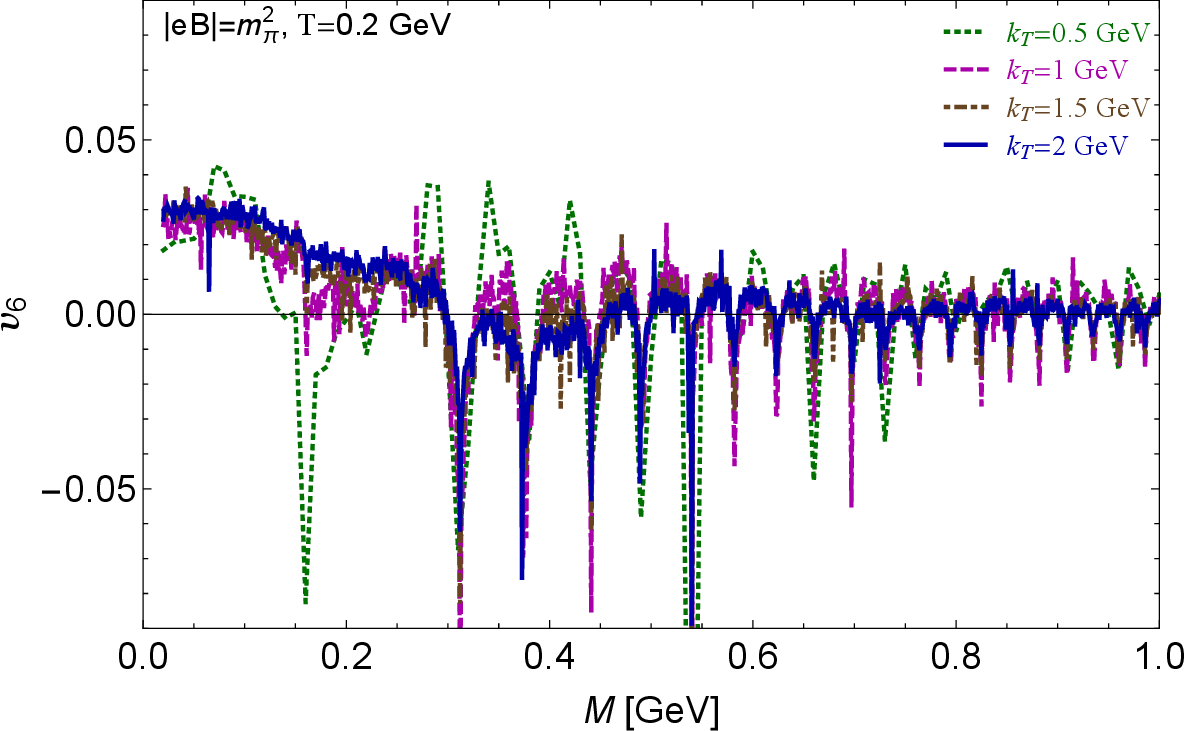}}
  \hspace{0.03\textwidth}
{\includegraphics[width=0.45\textwidth]{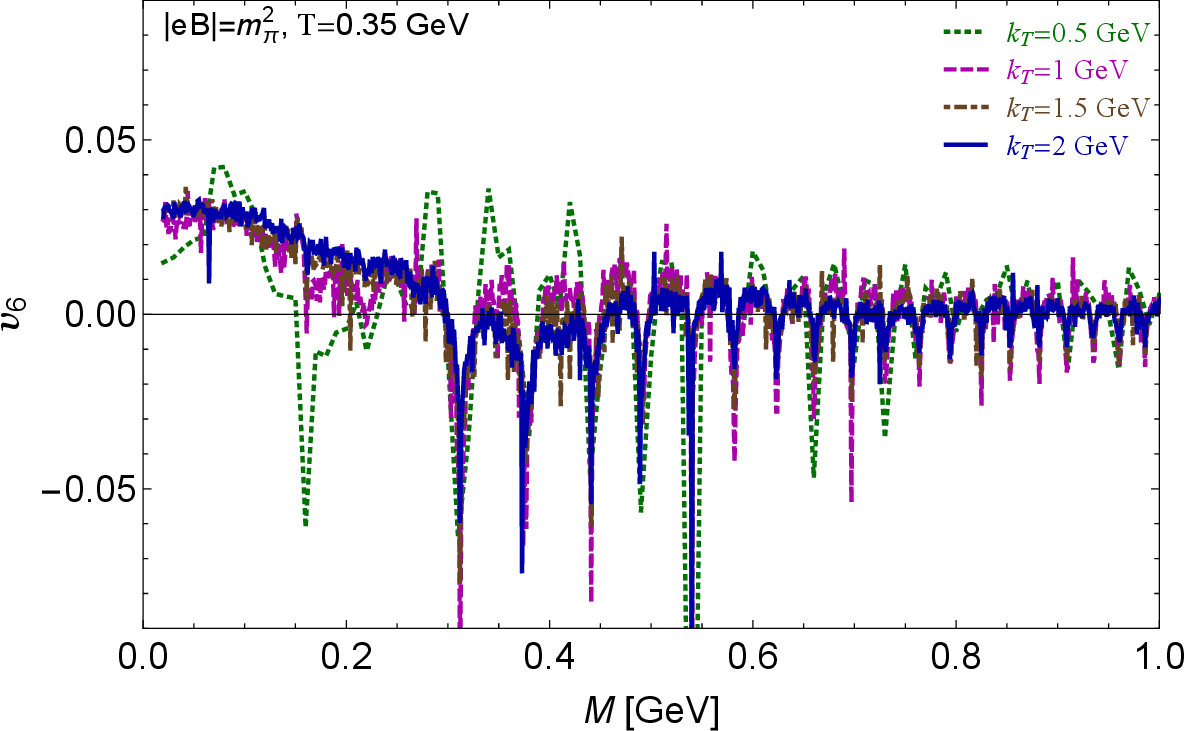}}\\[10pt]
{\includegraphics[width=0.45\textwidth]{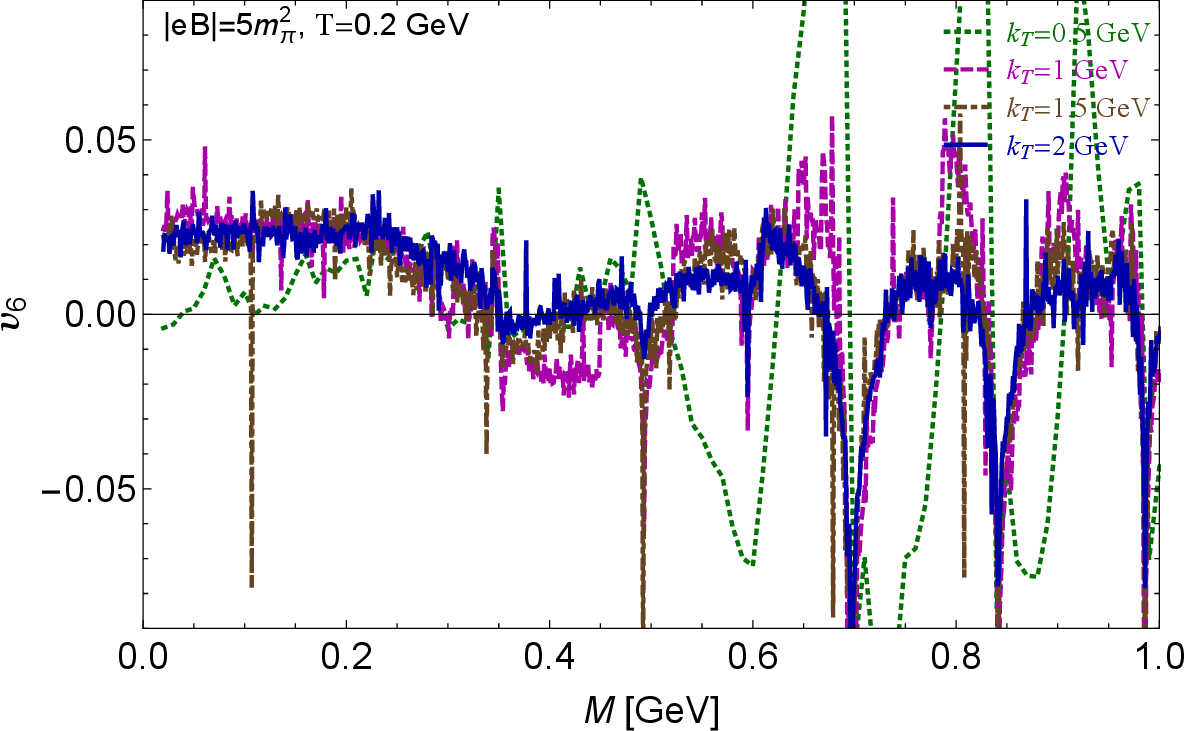}}
  \hspace{0.03\textwidth}
{\includegraphics[width=0.45\textwidth]{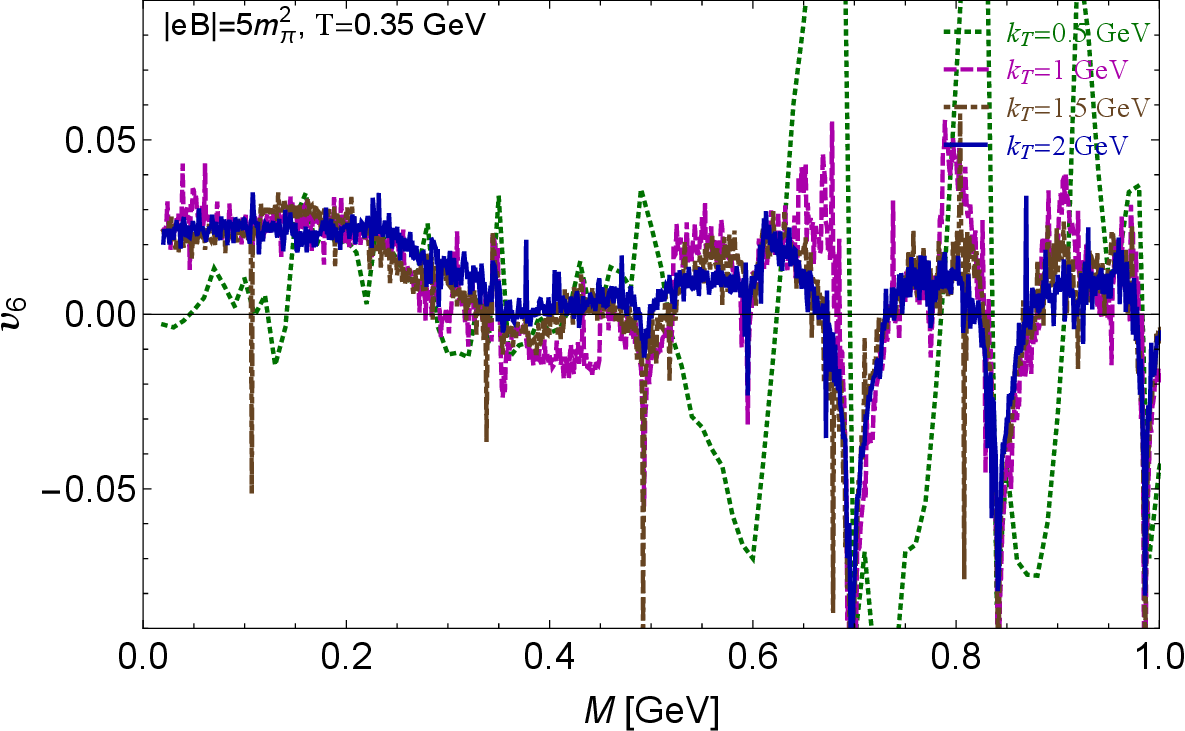}}\\
\caption{Anisotropic coefficient $v_6$ for the dilepton emission as a function of the invariant mass $M$ 
for several fixed values of the transverse momentum $k_T$. The top two panels correspond to $|eB|=m_\pi^2$, and the bottom two panels to $|eB|=5m_\pi^2$. The panels on the left are for $T=0.2~\mbox{GeV}$, and the ones on the right are for $T=0.35~\mbox{GeV}$.}
\label{fig:dilepton_v6}
\end{figure}

The dilepton results for $v_6$ are shown in Fig.~\ref{fig:dilepton_v6}. The four panels show the results for two temperatures, i.e., $T=0.2~\mbox{GeV}$ (two left panels) and $T=0.35~\mbox{GeV}$ (two right panels), and two magnetic fields $|eB|=m_\pi^2$ (two top panels) and $|eB|=5m_\pi^2$ (two bottom panels).  As we see, the anisotropy coefficient $v_6$ tends to be positive at small $M$, with the maximal values of the order of $0.02$. Such values are comparable to those of photon $v_6$ in Fig.~\ref{fig:photon_v6}.

It should be noted that nonvanishing $v_4$ and $v_6$ are barely resolved for the intermediate transverse momentum $k_T=0.5~\mbox{GeV}$, especially in the case of the stronger field $|eB|=5m_\pi^2$. It is not surprising as the corresponding $k_T$ is comparable to $\sqrt{|eB|}$. Nevertheless, the trend becomes unambiguous for the larger values of $k_T$. As anticipated, both $v_4$ and $v_6$ have a weak temperature dependence at sufficiently large transverse momenta. The key features of the dilepton anisotropy are summarized in Table~\ref{tab:anisotropy}.

In addition to the alternating sign pattern and the hierarchy of coefficients $v_n\propto 1/n^2$ in the region of small invariant masses, we can also identify other interesting features in the high-resolution data obtained here. For example, we see well-pronounced modulations in the $v_n$ dependence on the invariant mass. Indeed, by comparing the results in Figs.~\ref{fig:dilepton_v2} through \ref{fig:dilepton_v6}, one can easily identify correlated patterns of peaks in all anisotropy coefficients $v_n$. They are visible even in the region of moderately large invariant masses. As is easy to verify, they come from the Landau-level quantization of quarks. In heavy-ion physics, such modulations could be hard, if not impossible, to observe. Perhaps, they could have some phenomenological implications in other contexts.

\subsection{Application to heavy-ion collisions}
\label{heavy-ion-collisions}

The main goal of our study is to characterize spatial profiles of the photon and dilepton emission in the rest frame of a strongly magnetized plasma. We found that both emission types could be highly anisotropic. This finding implies that a background magnetic field serves as an additional (``intrinsic") source of anisotropy unrelated to the hydrodynamic flow of the plasma. Therefore, it is natural to suggest that the anisotropy coefficients observed in heavy-ion collisions should contain the following two distinct contributions:
\begin{equation}
v^{\rm obs}_n = v^{\rm flow}_n \oplus v^{\rm mag}_n,
\label{total-v_n}
\end{equation}
where $v^{\rm flow}_n$ is the usual hydrodynamic part while $v^{\rm mag}_n$ is an intrinsic part due to a nonzero magnetic field. The proxy for the latter is given by our analysis of the photon and dilepton emission in the rest frame of a magnetized plasma above. It should be noted that different types of anisotropy contributions are not necessarily independent or additive.

To perform a systematic study of quantifying and separating the two contributions in Eq.~(\ref{total-v_n}) in the context of heavy-ion collisions, one would require detailed numerical investigations. Possible phenomenological approaches include hydrodynamic simulations or molecular dynamics models that take into account the space-time evolution of the plasma. The corresponding studies in the presence of a magnetic field background have not been done yet. Such studies are also beyond the scope of this paper. 

Without detailed simulations, here we can give only qualitative arguments, supporting the general idea of an additional intrinsic source of anisotropy due to the background magnetic field that have been ignored before. We can also speculate that the corresponding anisotropy contribution could be substantial if the magnetic field is as strong as suggested by some estimates \cite{Skokov:2009qp,Voronyuk:2011jd,Deng:2012pc,Bloczynski:2012en,Tuchin:2015oka,Guo:2019mgh}. It could be also important that the magnetic field is particularly strong during the early stages of the plasma evolution, when hydrodynamic flow did not develop fully.

\section{Summary and Conclusions}
\label{sec:summary}

In this paper, we investigated the higher-order anisotropy coefficients $v_4$ and $v_6$ for photon and dilepton emission from a magnetized hot QGP in the rest frame. For both processes, we revealed several characteristic features in the dependence of the anisotropy coefficients on the kinematic parameters. The summary of the overall magnitudes and signs of the anisotropy coefficients is given in Table~\ref{tab:anisotropy}. 

In the case of photon emission, we find qualitatively different anisotropy patterns at small and large transverse momenta. At small momenta (i.e., $k_T\lesssim \sqrt{|eB|}$), the signs and absolute values of the anisotropy coefficients are as follows: $v_4\simeq +0.05$ and $v_6 \simeq -0.02$. At large momenta (i.e., $k_T\gtrsim \sqrt{|eB|}$), the signs of $v_n$ reverse, but the absolute values remain about the same, i.e., $v_4\simeq -0.05$ and $v_6 \simeq +0.02$. Combining these findings with the $v_2$ results in Ref.~\cite{Wang:2020dsr}, we see that the signs of even coefficients $v_{n}$ alternate. The absolute values gradually decrease with increasing $n$ in each kinematic region. Quantitatively, the scaling appears to go as $v_n\propto 1/n^2$

We find that the dilepton emission also has a noticeable anisotropy. However, it is well pronounced only in the kinematic regime with large transverse momenta (i.e., $k_T\gtrsim \sqrt{|eB|}$) and small invariant masses (i.e., $M\lesssim \sqrt{|eB|}$). The  signs and absolute values of the anisotropy coefficients are as follows: $v_4\simeq -0.05$ and $v_6 \simeq +0.02$. Supplementing these findings with the results for $v_2$ obtained in Ref.~\cite{Wang:2022jxx}, we see that the signs of even coefficients $v_{n}$ alternate, and their  absolute values decrease with increasing $n$. The quantitative scaling is similar to that in the photon emission. 

In application to QGP produced by noncentral heavy-ion collisions, one may argue that the magnetic field could be too weak, e.g., well below the scale set by the pion mass, to have observable effects. It is possible and, perhaps, even likely that the field is weak indeed. Nevertheless, we argue that even weak fields can affect the anisotropy of both photon and dilepton emissions in certain kinematic regions. Indeed, as we see from our calculations, the anisotropy is sizable even for the transverse momenta that are much larger than the magnetic field scale. This is analogous to the anisotropy of the classical synchrotron radiation. Admittedly, the effects on the photon emission may be diluted by the subleading gluon-mediated processes. Hopefully, the anisotropy does not vanish completely and could remain observable. The situation with dileptons might be better, however.  Indeed, the same leading-order diagrams contribute in the case with and without the background magnetic field. 

It is tempting to suggest that a set of the first few nonzero anisotropy coefficients $v_n$, extracted from the photon and dilepton data, can provide a distinctive fingerprint of the background magnetic field in a hot QGP produced by heavy-ion collisions. The current data with overwhelming background effects may not allow one to test this idea easily in experiment. Additionally, the task is complicated by the convolution with other sources of anisotropy such as hydrodynamics flow and initial state fluctuations. Nevertheless, we find it valuable to have concrete theoretical predictions for the intrinsic $v_n^{\rm mag}$ produced by the background magnetic field. The advances in experimental techniques, collision simulations, and data analysis could make the current hopeless task possible in the future.

To give reliable theoretical predictions for the observable signatures in heavy-ion experiments, one needs to combine the results of this study with realistic space-time models of QGP with expansion and nonuniform profiles. The latter requires the use of phenomenological models, for example, such as $3+1$ viscous hydrodynamic simulations in Refs.~\cite{Dion:2011pp,Schenke:2010rr}. The corresponding task is beyond the scope of this paper. It has to be undertaken, however, before one can reach the final conclusions about the emission anisotropy as a likely signature of a background magnetic field in heavy-ion collisions.

While the motivation of this study was triggered by potential applications in heavy-ion physics, it is instructive to mention that our main results may also find applications in astrophysics, where relativistic QED plasmas are common. With a suitable rescaling of the model parameters, our analysis can be easily generalized to QED plasmas under conditions in magnetars \cite{Kaspi:2017fwg}, supernovae \cite{Hardy:2000gg}, and gamma-ray bursts~\cite{Granot:2015xba}. It is reasonable to assume that the anisotropy profiles of the photon and dilepton emission will be similar.

\acknowledgements
The work of X. W. was supported by National Science Foundation of China under Grant No.~12147103.
The work of I. A. S. was supported by the U.S. National Science Foundation under Grant No.~PHY-2209470. 

\appendix

\section{Imaginary part of the Lorentz-contracted polarization tensor}
\label{sec:Polarization}

For convenience, here we quote the expression for the imaginary part of the Lorentz-contracted polarization tensor that appears in the photon and dilepton rates; see Eqs.~(\ref{diff-rate-photon}) and (\ref{rate_dilepton}), respectively. In the Landau-level representation, the corresponding analytical expression takes the  following form~\cite{Wang:2020dsr,Wang:2021ebh,Wang:2022jxx}:
\begin{eqnarray}
\mbox{Im} \left[\Pi^{\mu}_{R,\mu}\left(\Omega, \mathbf{k}\right)\right] &=&
\sum_{f=u, d} \frac{N_c\alpha_f}{2\pi  \ell_f^4} \sum_{n>n^\prime}^{\infty}  
\frac{g(n, n^{\prime}) 
\left\{
\Theta\left[( k_{-}^{f} )^2 - k_{\parallel}^2 \right]
-\Theta\left[ k_{\parallel}^2-(k_{+}^{f})^2\right] \right\}
 }{\sqrt{ \left[( k_{-}^{f} )^2 - k_{\parallel}^2 \right]\left[ (k_{+}^{f})^2- k_{\parallel}^2\right] } }
 \mathcal{F}_{n, n^{\prime}}^f(\xi_f)  
\nonumber\\
&-&\sum_{f=u, d} \frac{N_c\alpha_f}{4\pi  \ell_f^4} \sum_{n=0}^{\infty} 
\frac{g_0(n)\Theta\left[ k_{\parallel}^2-(k_{+}^{f})^2\right]}{\sqrt{ k_{\parallel}^2 \left[ k_{\parallel}^2-(k_{+}^{f})^2\right]} }
\mathcal{F}_{n, n}^f(\xi_f)    ,
\label{Im-Pi-final}
\end{eqnarray}
where $\Theta\left(x\right)$ is the Heaviside step function, $\alpha_f = e_f^2/(4\pi)$, $e_f$ is the flavor-dependent electric charge of the quark, $k_{\parallel}^2\equiv \Omega^2-k_z^2$,  $\xi_f = k_{\perp}^2\ell_{f}^{2}/2$ and $\ell_{f}=1/\sqrt{|e_fB|}$ is a flavor-dependent magnetic length. The Landau-level thresholds are determined by the following two transverse momenta:
\begin{equation}
k_{\pm}^{f} = \left|\sqrt{m^2+2n|e_f B|}\pm\sqrt{m^2+2n^{\prime}|e_f B|}\right|.
\end{equation}
Functions $g(n, n^{\prime})$ and $g_0(n)$ are determined by the quark distribution functions. In thermal equilibrium, they are given by
\begin{eqnarray}
g(n, n^{\prime}) &=& 2-\sum_{s_1,s_2=\pm}
n_F\left(\frac{\Omega}{2}  +s_1 \frac{\Omega(n-n^{\prime})|e_fB|}{k_{\parallel}^2}+s_2 \frac{|k_z|}{2k_{\parallel}^2}\sqrt{ \left(k_{\parallel}^2-(k_{-}^{f})^2 \right)\left( k_{\parallel}^2- (k_{+}^{f})^2\right)} \right), 
\label{gnn}  \\
g_0(n) &=& g(n, n)  =2 - 2\sum_{s=\pm}
n_F\left(\frac{\Omega}{2} +s \frac{|k_z|}{2|k_{\parallel}|}\sqrt{ k_{\parallel}^2 - 4(m^2+2n|e_f B|)}
\right) ,
\label{g0n}
\end{eqnarray}
where $n_F(E) = 1/\left[\exp\left(E/T\right)+1\right]$ is the Fermi-Dirac distribution function.
Finally, $\mathcal{F}_{n, n^{\prime}}^f(\xi) $ is the following flavor-dependent function of the transverse momentum:
\begin{equation}
\mathcal{F}^f_{n, n^{\prime}} (\xi)
= 8\pi \left(n+n^{\prime}+m^2\ell_f^2\right)\left[\mathcal{I}_{0}^{n,n^{\prime}}(\xi)+\mathcal{I}_{0}^{n-1,n^{\prime}-1}(\xi) \right]
+8\pi  \left(\frac{k_{\parallel}^2 -k_{\perp}^2}{2} \ell_f^2  -(n+n^{\prime})\right)
\left[\mathcal{I}_{0}^{n,n^{\prime}-1}(\xi)+\mathcal{I}_{0}^{n-1,n^{\prime}}(\xi) \right],
\end{equation}
and function $\mathcal{I}_{0}^{n,n^{\prime}}(\xi)$ is defined in terms of the Laguerre polynomials, i.e., 
\begin{equation}
\mathcal{I}_{0}^{n,n^{\prime}}(\xi) = \frac{(n^\prime)!}{n!} e^{-\xi}  \xi^{n-n^\prime}
\left(L_{n^\prime}^{n-n^\prime}\left(\xi\right)\right)^2
= \frac{n!}{(n^\prime)!}e^{-\xi} \xi^{n^\prime-n}
\left(L_{n}^{n^\prime-n}\left(\xi\right)\right)^2 .
\label{I0f-LL-form1} 
\end{equation}
Note that the two different representations for $\mathcal{I}_{0}^{n,n^{\prime}}(\xi)$ are equivalent. Note that, by definition, Laguerre polynomials with negative lower indices vanish.
 
\bibliographystyle{apsrev4-2}

\end{document}